\documentclass[preprint,12pt]{elsarticle}
\usepackage{amsmath}
\usepackage{amssymb}
\usepackage{multirow}
\usepackage{enumerate}
\usepackage{cases}
 \usepackage{amsthm}
 \usepackage{framed}
\usepackage{tabularx}
\usepackage{mathrsfs}

\newtheorem{Theorem}{Theorem} 
\newtheorem{Definition}{Definition}

\usepackage{algorithm}%
\usepackage{tabularx}
\usepackage{listings}%
\usepackage{caption}
\usepackage{mathtools} 
\usepackage{algpseudocode}
\usepackage{graphicx}
\usepackage{slashed}

\makeatletter
\newcommand{\multiline}[1]{%
  \begin{tabularx}{\dimexpr\linewidth-\ALG@thistlm}[t]{@{}X@{}}
    #1
  \end{tabularx}
}
\makeatother

\begin{document}

\begin{frontmatter}
\title{A Quantum Automatic Tool for Finding Impossible Differentials}


\author{Huiqin Xie$^{1,2}$\corref{1}}
\author{Qiqing Xia $^{3}$}
\author{Ke Wang$^{1}$}
\author{Yanjun Li$^{4}$}
\author{Li Yang$^{3}$}
\cortext[1]{Corresponding author email: xiehuiqindky@163.com}
\address{1. Department of Cryptography Science and Technology, Beijing Electronic Science and Technology Institute, Beijing 100070, China\\
2. Key Laboratory of Cryptography of Zhejiang Province, Hangzhou Normal University, Hangzhou 311121, China\\
3. Institute  of  Information  Engineering, Chinese Academy of Sciences, Beijing 100085, China\\
4. Information Industry Information Security Evaluation Center, The 15th Research Institute of China Electronics Technology Group Corporation, Beijing 100083, China}

\begin{abstract}
Due to the superiority of quantum computing, traditional cryptography is facing severe threat. This makes the security evaluation of cryptographic systems in quantum attack models significant and urgent. For symmetric ciphers, the security analysis heavily relies on cyptanalytic tools. Thus exploring the use of quantum algorithms to traditional cyptanalytic tools has drawn a lot of attention. In this study, we utilize quantum algorithms to improve impossible differential attack, and design two quantum automatic tools for searching impossible differentials. The proposed quantum algorithms exploit the idea of miss-in-the-middle and the properties of truncated differentials. We rigorously prove their validity and calculate the quantum resources required to implement them. Compared to existing classical automatic cryptanalysis, the quantum tools proposed have the advantage of accurately characterizing S-boxes while only requiring polynomial complexity, and can take into consideration the impact of the key schedules in single-key model.

\end{abstract}


\end{frontmatter}


Recently, the development of quantum computers has made steady and rapid progress. As soon as quantum computers are successfully built, traditional cryptography will be severely threatened. Utilizing Shor’s algorithm \cite{ref-proceeding1}, the adversaries possessing quantum computers are able to break the public key cryptosystems built on integer factorization problem, for instance, the RSA scheme widely used in secure communication. Besides public key cryptography, study on cryptanalysis of symmetric cryptography against quantum adversaries has also achieved many outstanding results. By Grover’s algorithm one can achieve a quadratic speedup when searching unordered databases \cite{ref-proceeding2}. Therefore, to restore the same ideal security with that in classical setting, the key lengths of symmetric ciphers must be doubled in quantum setting. 

The exhaustive attack can only evaluate the ideal security margin of cryptographic schemes. To accurately grasp the quantum security of currently used symmetric schemes, we also need to investigate other possible quantum attacks. In this direction Simon's algorithm \cite{ref-journal1} is frequently used. Kuwakado and Morri first applied Simon's algorithm to attack Feistel structure, and proposed a 3-round quantum distinguisher \cite{ref-proceeding3}. Then they also attacked Even-Mansour cipher using similar idea and successfully recovered the key \cite{ref-proceeding4}. Authors in \cite{ref-journal2} forged messages of the CBC-MAC scheme using the method showed in \cite{ref-proceeding3}. Kaplan \textit{et al}. then also further developed the result in \cite{ref-proceeding3} and attacked several symmetric systems including GCM, PMAC, CLOC and so on \cite{ref-proceeding5}, . Both \cite{ref-journal2} and \cite{ref-proceeding5} proved the correctness of quantum distinguisher even if the Feistel structure has round functions that are not permutations.

Leander and Alexander embedded Simon’s algorithm in Grover’s algorithm in order to find the correct key of FX structure \cite{ref-proceeding6}. Following this attack strategy, Dong and Wang broke Feistel schemes and got the key via the quantum distinguisher showed in \cite{ref-proceeding3}. Afterwards they applied same strategy to extract the key of the generalized Feistel ciphers  \cite{ref-journal3,ref-journal4}. The above attacks are all implemented under the quantum version of chosen-plaintext model, also called $Q_2$ model \cite{ref-proceeding7,ref-proceeding8,ref-proceeding9}. In this attack model the cryptographic oracle can be queried with superposition states. Authors in \cite{ref-journal5} studied quantum related-key notion \cite{ref-journal5} where quantum adversaries can use superposition states of related keys to query oracles. Hosoyamada \textit{et al}. then further investigated quantum related-key notion and recovered the key of 2-round Even-Mansour algorithm \cite{ref-journal6}. Jaques \textit{et al}. analyzed the complexity of Grover’s algorithm when attacking AES \cite{ref-proceeding10}.

Besides specific quantum attacks, it is also essential to study quantum versions of ctyptanalytic tools, like differential, integral and linear analysis. Zhou \textit{et al}. first started this direction and utilized Grover's algorithm to differential attack \cite{ref-journal7}. Kaplan \textit{et al}. subsequently  used Grover's algorithm to enhance some variants of differential attack and linear attack \cite{ref-proceeding10}.  Xie \textit{et al}. made use of Bernstein-Vazirani algorithm to search for high-probability differentials \cite{ref-journal7}. Authors in \cite{ref-proceeding11} implemented quantum collision attacks on Whirlpool and AES-MMO schemes via differential characteristics. Dong \textit{et al}. enhanced truncated differential analysis through quantum algorithms and broke Gr$\slashed{o}$stl-512 scheme and AES-MMO cipher \cite{ref-proceeding12}. 

\textbf{Our contributions.} In this article, we study the applications of Simon’s algorithm to cryptanalytic tools of symmetric ciphers. We bring the superiority of quantum computing into traditional impossible differential analysis, and design quantum automatic tools for searching impossible differentials. We first propose a basic quantum algorithm that can find impossible differentials by imitating the classical impossible differential technique. Then by allowing the differentials to be truncated, we present another improved quantum algorithm.  We provide the correctness proof of the proposed algorithms and evaluate their quantum complexity. Proposed quantum tools have the following advantages:

\begin{enumerate}[\textbullet]
\item The quantum algorithms can be implemented in Q1 attack model, without any query to the quantum encryption or decryption oracle. In contrast, many other quantum attacks \cite{ref-proceeding3,ref-proceeding4,ref-journal2,ref-proceeding5,ref-journal3,ref-journal4} require the adversaries to perform quantum queries with superposition states. Our quantum tools are much easier to realize and thus more practical.
\vskip 0.2cm

\item Classical automatic impossible differential cryptanalysis tools, such as the UID-tool \cite{ref-journal9}, the U-tool \cite{ref-proceeding13}, the WW-tool \cite{ref-proceeding14} and the MILP tool \cite{ref-proceeding15}, either cannot involve the internal constructions of S-boxes or can only characterize small-scale S-boxes. Especially when dealing with common 8-bit S-boxes, they can attack only very few rounds. Our quantum automatic tools fully leverage the parallel advantages of quantum computing, allowing for complete characterization of S-boxes while maintaining complexity within polynomial time.
\vskip 0.2cm

\item Classical automatic impossible differential cryptanalysis cannot take the key schedule into account in single-key model, while our tools includes the key schedule part when implementing the quantum circuit of encryption, which solves this problem. 
\end{enumerate}

\section{Preliminaries}
We present a simple overview of the necessary
concepts and related results. 

\subsection{Quantum Attack Models}

$n,m$ are two arbitrary positive integers. $F:\mathbb{F}_2^n\rightarrow\mathbb{F}_2^m$ is a Boolean function. If the unitary operation
\begin{equation}
U_F: \sum_{x,y}|x\rangle|y\rangle\rightarrow\sum_{x,y}|x\rangle|y\oplus F(x)\rangle,
\end{equation}
is realized by a quantum circuit, we say that this circuit evaluates $F$ quantumly. Any vectorial Boolean function can be evaluated by a quantum circuit constructed with gates in a finite but universal set of unitary gates. This kind of set is referred to as a universal gate set \cite{ref-book1}. For example, the Phase gate $S$, Hadamard gate $H$, non-clifford gate $T$ and controlled-NOT quantum gate $CNOT$ forms a universal gate set. Each gate in this set is calculated as a single operation. For any vectorial Boolean function $F$, we let the notation $|U_F|$ denote the amount of quantum universal gates required to implement $U_F$. 

When analyzing the quantum security of cryptographic primitives, two common attack models for adversaries are considered. One is Q1 attack model, where the adversaries can utilize quantum computers to do offline computations but can only make classical online queries. Another one is Q2 attack model, where the adversaries can execute quantum queries in addition. Namely, a Q2 adversary can also make query to cryptographic primitives with inputs in superposition states and get the quantum states of their outputs. Q2 attack model is more strict on the adversaries’ ability since querying the quantum oracles of cryptographic systems is usually not easy to realize in practice.

\subsection{Simon's Algorithm}
Given $F:\mathbb{F}_2^n\rightarrow\mathbb{F}_2^m$ and a private vector $s\in\mathbb{F}_2^n$ satisfying
$$
[F(x_1)=F(x_2)]\Leftrightarrow [x_1\oplus x_2\in\{0^n,s\}],\,\,\,\forall x_1,x_2\in \mathbb{F}_2^n,
$$
Simon's algorithm \cite{ref-journal1} is originally used to solve the period $s$. If a function has a such period, we say that this function meets Simon's promise. Finding $s$ consumes at least $O(2^{n/2})$ classical queries when using classical algorithms, but Simon's algorithm only consumes $O(n)$ quantum queries. With the quantum circuit of $F$, Simon's algorithm needs to repeat the steps below:
\begin{enumerate}
\item Prepare a $(n+m)$-qubit quantum state $|0^n\rangle|0^m\rangle$. Apply Hadamard transform $H^{\otimes n}$ to the left register, obtaining
$$
\frac{1}{\sqrt{2^n}}\sum_{x\in \mathbb{F}_2^n}|x\rangle|0^m\rangle.
$$

\item Implement the unitary operator $U_F$ of $F$, giving the state
$$
\frac{1}{\sqrt{2^n}}\sum_{x\in \mathbb{F}_2^n}|x\rangle|F(x)\rangle.
$$

\item Measure the last register to obtain a vector $F(z)$, then the rest register will be
$$
\frac{1}{\sqrt{2}}(|z\rangle+|z\oplus s\rangle).
$$

\item Perform Hadamard operators $H^{\otimes n}$ on the above state, getting
$$
\frac{1}{\sqrt{2^{n+1}}}\sum_{\gamma\in\mathbb{F}_2^n}(-1)^{\gamma\cdot z}[1+(-1)^{\gamma\cdot s}]|\gamma\rangle.
$$

\item Measure this state. If a vector $\gamma$ satisfies $\gamma\cdot s=1$, its amplitude must be 0. The measurement result $\gamma$ always satisfies $\gamma\cdot s=0$.
\end{enumerate}

The process of Simon's algorithm is to repeat the above subroutine $O(n)$ times, giving $n-1$ vectors that are perpendicular to $s$ and linear independent. Using linear algebra knowledge we can compute $s$ easily.

\begin{figure}[H]
\centering
\includegraphics[width=10cm]{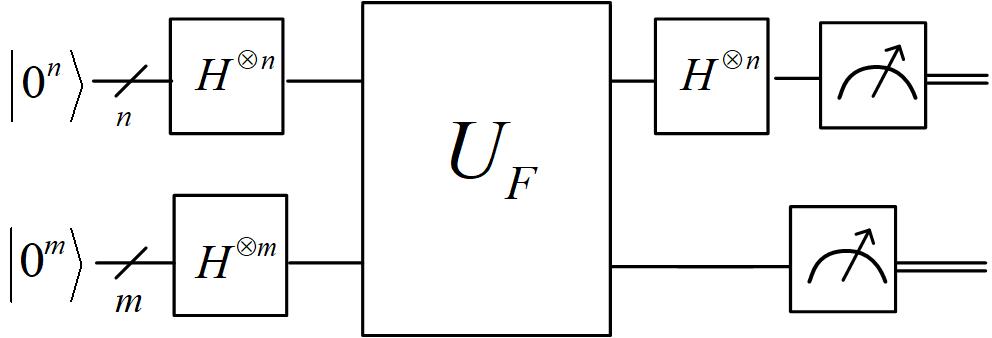}
\caption{Circuit illustration of Simon's algorithm\label{Fig1}}
\end{figure} 

We show quantum circuit illustration of the Simon's subroutine (steps 1-5) in Figure~\ref{Fig1}. Running steps 1-5 requires $2n$ Hadamard operators and 1 execution of the unitary operator $U_F$. Therefore, there are $m+n$ qubits and $O(2n^2+n|U_F|)$ gates totally in Simon's algorithm when run on $F$.

In cryptanalysis scenario, it is not always easy to construct a Boolean function that satisfies Simon’s promise. Even if a periodic function is constructed, there might be unwanted collisions that are not caused by the period. Kaplan \textit{et al}. relaxed Simon's promise and proved the following theorem.
\begin{Theorem}(\cite{ref-proceeding5})If $F:\mathbb{F}_2^n\rightarrow\mathbb{F}_2^n$ satisfies $\epsilon(F,s)\leq e_0<1$ for a period $s\in\mathbb{F}_2^n$ and some constant $e_0$, where
$$
\epsilon(F,s)=\max_{t\in\mathbb{F}_2^n\textbackslash\{0^n,s\}}\Pr_x[F(x\oplus t)=F(x)],
$$
then by repeating the subroutine $cn$ times, the probability of Simon's algorithm returning $s$ is no less than $1-(2(\frac{1+e_0}{2})^c)^n$.
\end{Theorem}

\subsection{Linear Structure}
We will transform the problem of seeking impossible differentials into a problem of seeking linear structures.
\begin{Definition} Given a function $F:\mathbb{F}_2^n\rightarrow\mathbb{F}_2^m$, $a\in\mathbb{F}_2^n$ is named a linear structure if
\begin{align}
F(x\oplus a)\oplus F(x)=b,\,\,\,\forall x\in \mathbb{F}_2^n
\end{align}
for some vector $b\in\mathbb{F}_2^m$. Namely, $F(x\oplus a)\oplus F(x)$ is constant.
\end{Definition}
For any $a\in\mathbb{F}_2^n$, $b\in\mathbb{F}_2^m$ satisfying Eq.(2), we call the pair $(a,b)$ as $F$'s linear structure duad. If $b$ is a zero vector $0^m$, then $a$ is called $F$'s period. If $(a_1,b_1)$, $(a_2,b_2)$ are two linear structure duads of $F$, then
$$
F(x\oplus a_1\oplus a_2)\oplus F(x)=F(x\oplus a_1)\oplus b_2\oplus F(x)=b_1\oplus b_2.
$$
Thus $(a_1,b_1)\oplus(a_2,b_2)$ is still $F$'s linear structure duad. All $F$'s linear structure duads form a subspace of the vector space $\mathbb{F}_2^{n+m}$. This subspace is named linear structure space and denoted by $L_F$.

For any vectors $v\in\mathbb{F}_2^m,u\in\mathbb{F}_2^n$, if there is $x\in\mathbb{F}_2^n$ satisfying $F(x)\oplus F(u\oplus x)=v$, then $(u,v)$ is said to make a ``\emph{match}'' of $F$ at $x$. Being linear structure duad is equivalent to causing \emph{matches} of $F$ at all points $x\in\mathbb{F}_2^n$. 

\section{A Basic Quantum Tool for Finding Impossible Differentials}
We present a universal quantum algorithm which finds impossible differentials of an
arbitrary block cipher. The main idea is to use probability-1 differentials to construct impossible differentials. Since the probability-1 differentials of an encryption function are also its linear structure duads, one can find them through constructing a quantum algorithm that finds linear structure duads. We first show a quantum algorithm that finds linear structure duads, then based on this algorithm we propose a basic quantum tool for impossible differentials.

\subsection{Finding Linear Structure Duads via Simon's Algorithm}
$L_F$ is the linear structure space of function $F:\mathbb{F}_2^n\rightarrow\mathbb{F}_2^m$ as defined in Section 2.3. Namely,
$$
L_F=\{(a,b)\in\mathbb{F}_2^{n}\times\mathbb{F}_2^{m}|F(x)\oplus F(x\oplus a)=b, \forall x\in\mathbb{F}_2^{n}\}.
$$
We aim to obtain $L_F$.  Here the value of $m$ does not need to be equal to $n$. Define a new function 
\begin{align}
G:\mathbb{F}_2^n\times\mathbb{F}_2^m&\rightarrow\mathbb{F}_2^m\notag\\
(x,y)\,\,\,&\rightarrow F(x)\oplus y.
\end{align}
For any duad $(a,b)\in\mathbb{F}_2^{n}\times\mathbb{F}_2^{m}$, if $(a,b)$ is $G$'s period, it will also be $F$'s linear structure duad. Therefore, $F$'s linear structure duads can be found by Simon's algorithm. Based on such analysis, we propose a algorithm named FindStruct as below, which is used to find linear structure duads.

\begin{algorithm}[H]
\caption*{{\large \textbf{Algorithm FindStruct}}}
\hangafter 1
\hangindent 4.1em
\textbf{Input$\,$}: $\,\,$a parameter $c$ and the access to the quantum unitary operator $U_F$ of a function $F:\mathbb{F}_2^n\rightarrow\mathbb{F}_2^m$.

\hangafter 1
\hangindent 4.7em
\textbf{Output}: the linear structure space $L_F$.

\begin{algorithmic}[1]
    \For{$i=1$ to $c(n+m)$}
        \State \parbox[t]{\dimexpr\textwidth-\leftmargin-\labelsep-\labelwidth}{Prepare a $(n+2m)$-qubit state $|0^n\rangle|0^m\rangle|0^m\rangle$ and implement the Hadamard gate $H^{\otimes(n+m)}$ to get $$|\Psi_1\rangle=\frac{1}{\sqrt{2^{n+m}}}\sum_{x\in\mathbb{F}_2^n, y\in\mathbb{F}_2^m}|x\rangle|y\rangle|0^m\rangle.$$\strut}
        
        \State \parbox[t]{\dimexpr\textwidth-\leftmargin-\labelsep-\labelwidth}{Use the unitary operator $U_F$ to get the state $$|\Psi_2\rangle=\frac{1}{\sqrt{2^{n+m}}}\sum_{x\in\mathbb{F}_2^n, y\in\mathbb{F}_2^m}|x\rangle|y\rangle|F(x)\rangle.$$\strut}
        \State \parbox[t]{\dimexpr\textwidth-\leftmargin-\labelsep-\labelwidth}{Apply CNOT operators to the last two registers to get the state$$|\Psi_3\rangle=\frac{1}{\sqrt{2^{n+m}}}\sum_{x\in\mathbb{F}_2^n, y\in\mathbb{F}_2^m}|x\rangle|y\rangle|F(x)\oplus y\rangle.$$\strut}
         \State \parbox[t]{\dimexpr\textwidth-\leftmargin-\labelsep-\labelwidth}{Measure the rightmost register to get a value $z\in\mathbb{F}_2^m$, then there exist vectors $x_0\in\mathbb{F}_2^n$, $y_0\in\mathbb{F}_2^m$ such that $F(x_0)\oplus y_0=z$. Thus the two leftmost registers are collapsed to
         \begin{equation}\frac{1}{\sqrt{|S_z|}}\sum_{(x,y)\in S_z}|x\rangle|y\rangle,\end{equation}
         where $S_z=\{(x,y)\in\mathbb{F}_2^{n+m}|F(x)\oplus y=z\}$.\strut}
\State \parbox[t]{\dimexpr\textwidth-\leftmargin-\labelsep-\labelwidth}{Implement the Hadamard gate $H^{\otimes(n+m)}$ on the above state to get
         $$\frac{1}{\sqrt{|S_z|2^{n+m}}}\sum_{ \gamma_1\in\mathbb{F}_2^n \atop \gamma_2\in \mathbb{F}_2^m}\sum_{(x,y)\in S_z}(-1)^{x\cdot \gamma_1\oplus y\cdot \gamma_2}|\gamma_1\rangle|\gamma_2\rangle,$$
         then measure this state to get a vector $\gamma^{(i)}\in\mathbb{F}_2^{n+m}$.\strut}
    \EndFor
\algstore{bkbreak}
\end{algorithmic}
\end{algorithm}

\begin{algorithm}
\begin{algorithmic}[1]
\algrestore{bkbreak}
    \State \parbox[t]{\dimexpr\textwidth-\leftmargin-\labelsep-\labelwidth}{After getting the vectors $\gamma^{(1)},\gamma^{(2)},\cdots,\gamma^{(c(m+n))}\in\mathbb{F}_2^{n+m}$ by steps 1-7, solve the following linear equation
    \begin{equation}\left\{\begin{array}{l}\gamma^{(1)}\cdot (x,y)=0\\\gamma^{(2)}\cdot (x,y)=0\\\vdots\\\gamma^{(c(m+n))}\cdot (x,y)=0,\\
    \end{array}\right.\end{equation}
    where $(x,y)\in\mathbb{F}_2^{n}\times\mathbb{F}_2^{m}$ are unknowns, and output its solution space.\strut}
    \end{algorithmic}
\end{algorithm}

\begin{figure}[H]
\centering
\includegraphics[width=10cm]{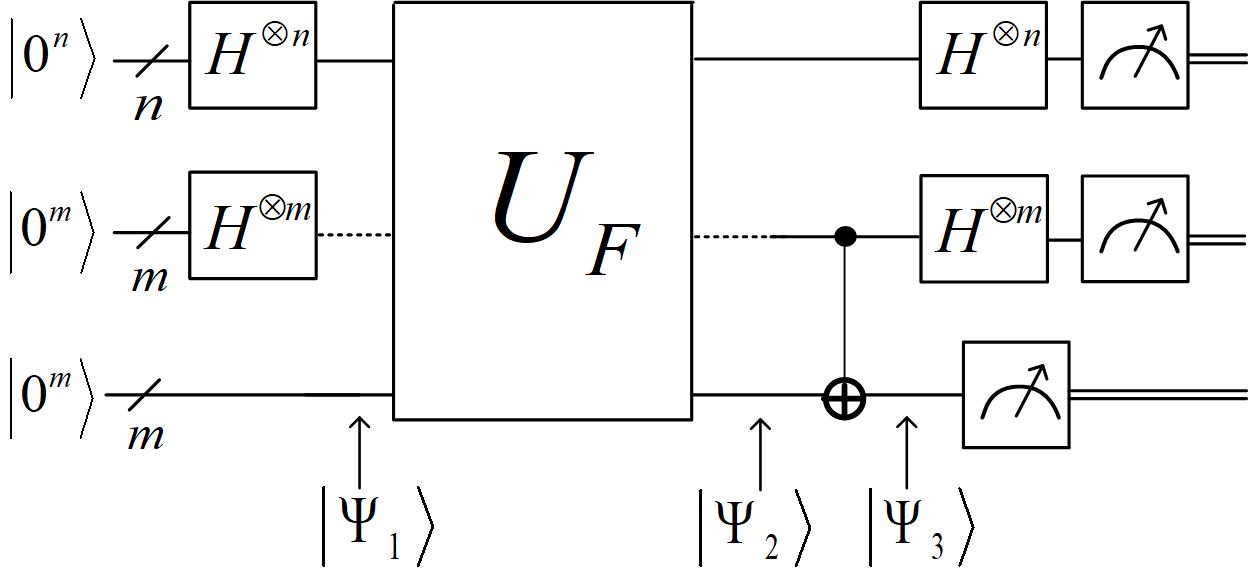}
\caption{Quantum circuit of FindStruct subroutine \label{Fig2}}
\end{figure}

The quantum computing part of algorithm FindStruct is to repeat steps 2-6 independently for $c(m+n)$ times. We call steps 2-6 as FindStruct subroutine. Its quantum circuit is shown in Figure~\ref{Fig2}. Steps 1-7 are actually to execute $c(m+n)$ times of Simon’s subroutine on $G(x,y)=F(x)\oplus y$ to independently obtain $c(m+n)$ vectors $\gamma^{(1)},\cdots,\gamma^{(c(m+n))}$. We expect that, as original Simon's algorithm, the periods of $G$ are orthogonal to $\gamma^{(1)},\cdots,\gamma^{(c(m+n))}$, thus we can get the periods of $G$ by solving Eq.(5), which are also linear structure duads of $F$. Theorem 1 gives the condition for Simon's algorithm to output the periods. Therefore, in order for algorithm FindStruct to successfully output $L_F$, function $G$ must satisfy this condition. However, $G$ may have more than one period since function $F$ may have more than one linear structure duads, or the length of $G$'s output is not equal to the length of input. Therefore, simply applying Theorem 1 is not sufficient to justify the soundness of algorithm FindStruct. To this end, we define a new parameter $\theta(\cdot)$. For function $F:\mathbb{F}_2^n\rightarrow\mathbb{F}_2^m$, defined
\begin{align}
\theta(F)=&\max_{a\in \mathbb{F}_2^n  b\in \mathbb{F}_2^m \atop (a,b)\notin L_F}{\rm \large Pr}_{x}\big[F(x)\oplus F(x\oplus a)=b\big]\notag\\
=&\max_{a\in \mathbb{F}_2^n  b\in \mathbb{F}_2^m \atop (a,b)\notin L_F}\frac{1}{2^n}\Big|\big{\{}x\in\mathbb{F}_2^n|F(x)\oplus F(x\oplus a)=b\big{\}}\Big|.
\end{align}
It is obvious that $0\leq\theta(F)<1$. If $(a,b)$ is in $L_F$, i.e., it is $F$'s linear structure duad, then it will cause a \emph{match} of $F$ at each point $x\in\mathbb{F}_2^n$. If $(a,b)\notin L_F$, then the number of \emph{matches} caused by $(a,b)$ will be less than $2^n$. The closer the value of $\theta(F)$ is to zero, the fewer \emph{matches} that the vector $(a,b)$ not in $L_F$ can cause. Theorem 2 shows the validity of algorithm FindStruct.

\begin{Theorem} Let $L$ be the solution set output by algorithm FindStruct run on $F:\mathbb{F}_2^n\rightarrow\mathbb{F}_2^m$ with a parameter $c$, then $L_F\subseteq L$. Moreover, if there is a constant $e_0$ such that $\theta(F)\leq e_0<1$, then the probability of $L_F$ being equal to $L$ is no less than $1-\big(2(\frac{1+e_0}{2})^c\big)^{n+m}$.
\end{Theorem}

The idea of proving Theorem 2 is almost the same to that of Theorem 1 in \cite{ref-proceeding5}, except that the case where the function has multiple periods or linear structures and the case the lengths of output and input are unequal need to be taken into account. We show the proof in the appendix.

According to Theorem 2, taking $c$ greater than $3/(1-e_0)$ can guarantee that the probability of algorithm FindStruct outputting vectors not in $L_F$ decreases exponentially with $n$. The condition $\theta(F)\leq e_0<1$ means that the vectors not linear structure duads of $F$ should not cause too much \emph{matches},  or in other words, vectors that are not periods of $G$ defined in Eq.(3) should not cause too much collisions. 

\subsection{Quantum Tool for Impossible Differentials}

The method of finding impossible differentials is to find probability-1 differential characteristics that propagate respectively from the input end and the output end of the cipher but can not match when they meet. 

\begin{figure}[H]
\centering
\includegraphics[width=5.5cm]{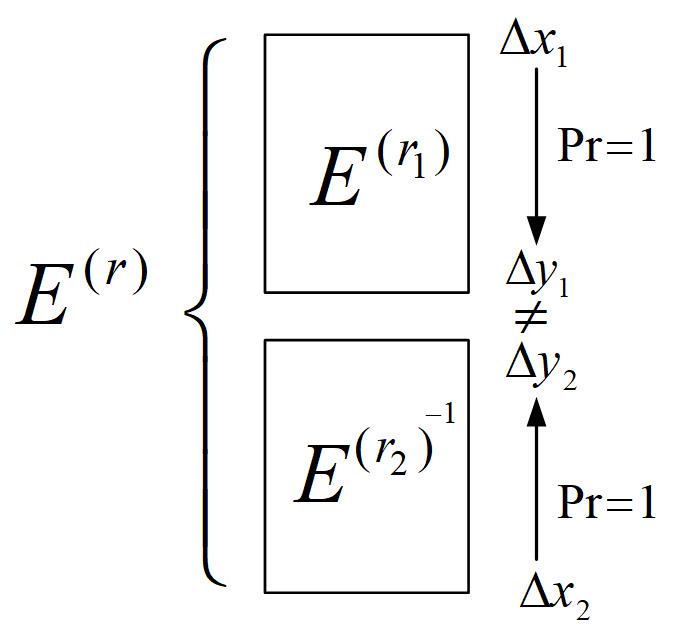}
\caption{Construction idea of impossible differentials\label{Fig3}}
\end{figure}

$E^{(r)}$ is an arbitrary block cipher that has $r$ rounds. $E$ denotes the round function. The blocksize is $n$ and key space is $\mathcal{K}=\mathbb{F}_2^m$. For each $k\in\mathcal{K}$, the output of $E^{(r)}$ on plaintext $x$ is $E_k^{(r)}(x)$. Our goal is to get impossible differentials of $E^{(r)}$. Namely, find $(\alpha,\beta)\in\mathbb{F}_2^n\times\mathbb{F}_2^n$ such that
$$
E_k^{(r)}(x)\oplus E_k^{(r)}(\alpha\oplus x)\neq \beta, \forall x\in\mathbb{F}_2^n, \forall k\in\mathbb{F}_2^m.
$$
We divide $E^{(r)}$ into two functions $E^{(r)}=E^{(r_2)}\circ E^{(r_1)}$. Here $1\leq r_1,r_2\leq r-1$ and $r_1+r_2=r$. Let ${E^{(r_2)}}^{-1}$ be the inverse function of $E^{(r_2)}$. As shown in Figure~\ref{Fig3}, if $(\Delta x_1,\Delta y_1)$ is a differential of $E^{(r_1)}$, $(\Delta x_2,\Delta y_2)$ is a differential of ${E^{(r_2)}}^{-1}$, satisfying
\begin{equation*}
\begin{split}
&E_{k}^{(r_1)}(x\oplus \Delta x_1)\oplus E_{k}^{(r_1)}(x)=\Delta y_1, \forall x\in\mathbb{F}_2^n, \forall k\in\mathbb{F}_2^m\\
&{E_{k}^{(r_2)}}^{-1}\!(x\oplus \Delta x_2)\oplus {E_{k}^{(r_2)}}^{-1}\!(x)=\Delta y_2, \forall x\in\mathbb{F}_2^n, \forall k\in\mathbb{F}_2^m,
\end{split}
\end{equation*}
and $\Delta y_1\neq\Delta y_2$, then $(\Delta x_1,\Delta x_2)$ will be an impossible differential of $E^{(r)}$. Therefore, to find impossible differentials of $E^{(r)}$, we only need to obtain differentials of $E^{(r_1)}$ and ${E^{(r_2)}}^{-1}$ with a probability of 1.

For any $t$-round block cipher $E^{(t)}$ that has key length $m$ and blocksize $n$, we treat both the plaintext and key as inputs of $E^{(t)}$, then the function
\begin{equation*}
\begin{split}
E^{(t)}:\mathbb{F}_2^m\times\mathbb{F}_2^n&\rightarrow\mathbb{F}_2^n\\
(k,x)&\rightarrow E_k^{(t)}(x)
\end{split}
\end{equation*}
is public and completely determined. The Q1 adversaries can construct the unitary operator 
$$
U_{E^{(t)}}:\!\!\sum_{(k,x)\in\mathbb{F}^{m+n}\atop y\in\mathbb{F}_2^n}\!\!|k,x\rangle|y\rangle\longrightarrow\sum_{(k,x)\in\mathbb{F}^{m+n}\atop y\in\mathbb{F}_2^n}\!\!|k,x\rangle|y\oplus E^{(t)}_k(x)\rangle
$$ 
by himself. Since
\begin{equation*}
\begin{split}
E^{(t)}(k\oplus 0^m,x\oplus\Delta x)\oplus E^{(t)}(k,x)=\Delta y, 
\forall (k,x)\in\mathbb{F}_2^{m+n}\\
\iff E_{k}^{(t)}(x\oplus \Delta x)\oplus E_{k}^{(t)}(x)=\Delta y, \forall k\in\mathbb{F}_2^m, \forall x\in\mathbb{F}_2^n, 
\end{split}
\end{equation*}
if $((0^m,\Delta x),\Delta y)$ is $E^{(t)}$'s linear structure duad, then $(\Delta x,\Delta y)$ will be $E^{(t)}$'s differential of probability 1. Thus one can use algorithm FindStruct to get $E^{(t)}$'s differentials of probability 1, with an additional requirement that the first $m$ bits of the linear structures should be 0. This can be achieved by adding additional equations to Eq.(5) when run algorithm FindStruct on $E^{(t)}$. By the above analysis, we design the following algorithm  for searching impossible differentials.
\begin{algorithm}[H]
\caption*{{\large \textbf{Algorithm FindImpoDiff}}}
\hangafter 1
\hangindent 3.7em
\textbf{Input$\,$}: $\,\,$a parameter $c$ and a cipher $E^{(r)}:\mathcal{K}\times\mathbb{F}_2^n\rightarrow\mathbb{F}_2^n$. ($\mathcal{K}=\mathbb{F}_2^m$ is the key space.)

\hangafter 1
\hangindent 3.7em
\textbf{Output}: Impossible differentials of $E^{(r)}$.

\begin{algorithmic}[1]
    \For{$r_1=1$ to $r-1$}
        \State \parbox[t]{\dimexpr\textwidth-\leftmargin-\labelsep-\labelwidth}{Let $r_2=r-r_1$, divide $E^{(r)}$ into two functions $E^{(r)}=E^{(r_2)}\circ E^{(r_1)}$.\strut}
        \State \parbox[t]{\dimexpr\textwidth-\leftmargin-\labelsep-\labelwidth}{Run steps 1-7 of algorithm FindStruct on $E^{(r_1)}:\mathbb{F}_2^{m}\times\mathbb{F}_2^n\rightarrow\mathbb{F}_2^n$ with parameter $c$ getting $c(m+2n)$ vectors $\gamma^{(1)},\gamma^{(2)},\cdots,\gamma^{c(m+2n)}\in\mathbb{F}_2^{m+2n}$.\strut}
        \State \parbox[t]{\dimexpr\textwidth-\leftmargin-\labelsep-\labelwidth}{Solve the equation
          \begin{equation}\left\{\begin{array}{l}\gamma^{(1)}\cdot(k,x,y)=0\\ \gamma^{(2)}\cdot(k,x,y)=0\\\vdots\\ \gamma^{(c(m+2n))}\cdot(k,x,y)=0\\
          k=0^{m},
          \end{array}\right.\end{equation}where $(k,x,y)\in\mathbb{F}_2^{m+n+n}$ are unknowns, to get the solution set $A_{r_1}$. \strut}
        \State \parbox[t]{\dimexpr\textwidth-\leftmargin-\labelsep-\labelwidth}{Run steps 1-7 of algorithm FindStruct on ${E^{(r_2)}}^{-1}:\mathbb{F}_2^{m}\times\mathbb{F}_2^n\rightarrow\mathbb{F}_2^n$ with parameter $c$ getting $c(m+2n)$ vectors $\tilde{\gamma}^{(1)},\tilde{\gamma}^{(2)},\cdots,\tilde{\gamma}^{c(m+2n)}\in\mathbb{F}_2^{m+2n}$.\strut}
        \State \parbox[t]{\dimexpr\textwidth-\leftmargin-\labelsep-\labelwidth}{Solve the equation
          \begin{equation}\left\{\begin{array}{l} \tilde{\gamma}^{(1)}\cdot(k,x,y)=0\\ \tilde{\gamma}^{(2)}\cdot(k,x,y)=0\\\vdots\\\tilde{\gamma}^{(c(m+2n))}\cdot(k,x,y)=0\\
          k=0^{m},
          \end{array}\right.\end{equation}where $(k,x,y)\in\mathbb{F}_2^{m+n+n}$ are unknowns, to get the solution set $B_{r_2}$. \strut} 
\algstore{bkbreak}
\end{algorithmic}
\end{algorithm}

\begin{algorithm}
\begin{algorithmic}[1]
\algrestore{bkbreak}
        \For{$(0^{m},\Delta x_1,\Delta y_1)\in A_{r_1}$}
        \For{$(0^{m},\Delta x_2,\Delta y_2)\in B_{r_2}$}
        \If{$\Delta x_1\neq 0^n$  $\wedge$ $\Delta x_2\neq 0^n$  $\wedge$  $\Delta y_1\neq \Delta y_2$}
        \State \parbox[t]{\dimexpr\textwidth-\leftmargin-\labelsep-\labelwidth}{Output $(\Delta x_1,\Delta x_2)$.\strut}
        \EndIf
         \EndFor
         \EndFor
    \EndFor
\end{algorithmic}
\end{algorithm}

Steps 3-4 are to find differentials of $E^{(r_1)}$. Steps 5-6 are to get differentials of ${E^{(r_2)}}^{-1}$. Since $E^{(r_1)}$ and ${E^{(r_2)}}^{-1}$ are public and determinate functions, the adversary can execute the unitary operators $U_{E^{(r_1)}}$ and $U_{{E^{(r_2)}}^{-1}}$ when calling FindStruct subroutine. 

Given a block cipher $E^{(r)}$, define
$$\Theta(E^{(r)})=\max\{\theta(E^{(t)})|1\leq t\leq r-1\},$$
where $E^{(t)}$ is $t$-round reduced version of $E^{(r)}$ and $\theta(E^{(t)})$ is defined as Eq.(6). That is,
\begin{equation*}
    \begin{split}        \theta(E^{(t)})=\max_{{(a_1,a_2)\in\mathbb{F}_2^{m}\times\mathbb{F}_2^{n}\atop b\in\mathbb{F}_2^{n}}\atop ((a_1,a_2),b)\notin L_{E^{(t)}}}\frac{1}{2^{m+n}}\Big|\big\{(k,x)\in\mathbb{F}_2^{m}\times\mathbb{F}_2^{n}|E^{(t)}(k,x)\oplus
    E^{(t)}(k\oplus a_1,x\oplus a_2)=b\big\}\Big|,
    \end{split}
\end{equation*}
where $L_{E^{(t)}}$ is the linear structure space of $E^{(t)}$. $\theta(E^{(t)})$ is the maximum amount of \emph{matches} that the vectors not linear structure duads of $E^{(t)}$ can cause. Therefore, the smaller the parameter $\Theta(E^{(r)})$ is, the fewer \emph{matches} the vectors not linear structure duads of the reduced version of $E^{(r)}$ can cause. According to Theorem 2, the following theorem holds.
\begin{Theorem} Block cipher $E^{(r)}$ satisfies $\Theta(E^{(r)})\leq e_0<1$ for a constant $e_0$. If executing algorithm FindImpoDiff on $E^{(r)}$ with parameter $c$ outputs $(\Delta x_1,\Delta x_2)$, the probability of $(\Delta x_1,\Delta x_2)$ being $E^{(r)}$'s impossible differential is no less than $1-2\big(2(\frac{1+e_0}{2})^c\big)^{2n+m}$, where
$m$ is the key length and $n$ is the blocksize.
\end{Theorem}

According to Theorem 3, taking $c$ greater than $3/(1-e_0)$ guarantees that the probability of algorithm FindImpoDiff outputting the vectors not being impossible differentials decreases exponentially with $n$.

It should be noted that, according to Theorem 2, any linear structure duad of $E^{(r_1)}$ whose first $m$ bits are zero must belong to the solution set $A_{r_1}$. On the other hand, $((0^{m},\Delta x_1),\Delta y_1)$ being a linear structure duad of $E^{(r_1)}$ is equivalent to $(\Delta x_1,\Delta y_1)$ being a probability-1 differential of $E^{(r_1)}$. Thus all probability-1 differentials of $E^{(r_1)}$ must be in the set $A_{r_1}$. Similarly, all probability-1 differentials of ${E^{(r_2)}}^{-1}$ must be in the set $B_{r_2}$. Therefore, all impossible differentials linked by two differentials of probability 1 as in Figure~\ref{Fig3} must be output by algorithm FindImpoDiff. This holds even if the condition $\Theta(E^{(r)})\leq e_0<1$ is not satisfied. The condition  $\Theta(E^{(r)})\leq e_0<1$ is only used to ensure that the probability of incorrectly outputting a vector that is not impossible differential is exponentially small.

\subsection{Complexity of Algorithm FindImpoDiff} 
The process of algorithm FindImpoDiff does not involve quantum queries and thus can be executed by Q1 adversaries. We evaluate the complexity by calculating the amount of qubits and quantum gates needed.

In algorithm FindImpoDiff, finding probability-1 differentials of $E^{(r_1)}$ requires to execute FindStruct subroutine on $E^{(r_1)}$ for $c(m+2n)$ times. Each execution includes $2m+4n$ Hadamard gates, $n$ CNOT gates and one unitary operator $U_{E^{(r_1)}}$. Finding probability-1 differentials of ${E^{(r_2)}}^{-1}$ requires to execute FindStruct subroutine on ${E^{(r_2)}}^{-1}$ for $c(m+2n)$ times. Each execution includes $2m+4n$ Hadamard gates, $n$ CNOT gates and one unitary operator $U_{{E^{(r_2)}}^{-1}}$. Thus, the amount of Hardamard gates in algorithm FindImpoDiff is
\begin{equation*}
    \begin{split} 
&\sum_{r_1=1,\cdots,r-1\atop r_2=r-r_1}[c(m+2n)(2m+4n)+c(2n+m)(4n+2m)]\\
&=\sum_{r_1=1}^{r-1}(4m+8n)c(2n+m)\\
&=4c(r-1)(m^2+4n^2+4nm)\in O(n^2).
    \end{split}
\end{equation*}
The amount of CNOT gates in algorithm FindImpoDiff is
\begin{equation*}
    \begin{split} 
&\sum_{r_1=1,\cdots,r-1\atop r_2=r-r_1}[(2n+m)cn+(2n+m)cn]\\
&=\sum_{r_1=1}^{r-1}(4cn^2+2cmn)\\
&=2(r-1)c(2n^2+mn)\in O(n^2).
    \end{split}
\end{equation*}
Algorithm FindImpoDiff also requires to execute the unitary operators $U_{E^{(r_1)}}$ and $U_{{E^{(r_2)}}^{-1}}$ $c(m+2n)$ times for each $1\leq r_1\leq r-1,r_2=r-r_1$. As explained in Section 2.1, $|U_{E^{(r_1)}}|$ and $|U_{{E^{(r_2)}}^{-1}}|$ denote the numbers of universal gates required to implement $U_{E^{(r_1)}}$ and $U_{{E^{(r_2)}}^{-1}}$, respectively. Since 
\begin{equation*}
    \begin{split} 
&\sum_{r_1=1,\cdots,r-1\atop r_2=r-r_1}c(m+2n)|U_{E^{(r_1)}}|+c(m+2n)|U_{{E^{(r_2)}}^{-1}}|\\
&= c(m+2n)\sum_{r_1=1,\cdots,r-1\atop r_2=r-r_1}(|U_{E^{(r_1)}}|+|U_{E^{(r_2)}}|)\\
&=c(m+2n)\sum_{r_1=1}^{r-1}|U_{E^{(r)}}|\\
&=(r-1)c(m+2n)|U_{E^{(r)}}|\in O(poly(n)),
    \end{split}
\end{equation*}
algorithm FindImpoDiff additionally requires to execute the quantum circuit of $E^{(r)}$ for $(r-1)c(m+2n)$ times. The quantum resources to implement algorithm FindImpoDiff are listed in Table 1.

Then we calculate the amount of qubits needed for algorithm FindImpoDiff. Running FindStruct subroutine on $E^{(r_1)}$ needs $(m+n)+n+n=m+3n$ qubits. Running FindStruct subroutine on ${E^{(r_2)}}^{-1}$ also needs $m+3n$ qubits. Due to the reusability of qubits, $m+3n$ qubits are enough to execute algorithm FindImpoDiff.

In addition to the quantum computing part, algorithm FindImpoDiff also needs to solve linear equations. Solving  Eq.$(7)$ is equivalent to solving the equation
\begin{equation*}\left\{\begin{array}{l} (\gamma^{(1)}_{m+1},\gamma^{(1)}_{m+2},\cdots,\gamma^{(1)}_{m+2n})\cdot (x,y)=0\\ (\gamma^{(2)}_{m+1},\gamma^{(2)}_{m+2},\cdots,\gamma^{(2)}_{m+2n})\cdot (x,y)=0\\\vdots\\ (\gamma^{(c(m+2n))}_{m+1},\gamma^{(c(m+2n))}_{m+2},\cdots,\gamma^{(c(m+2n))}_{m+2n})\cdot (x,y)=0.
          \end{array}\right.
\end{equation*}
Here $\gamma^{(j)}_{i}$ denotes the $i$-th bit of $\gamma^{(j)}$ ($1\leq j\leq c(m+2n)$). This linear system contains $2n$ unknowns and $c(m+2n)$ equations. The complexity of solving this linear system by Gaussian elimination is $O(cn^2(m+2n))$. Similarly, the complexity of solving Eq.$(8)$ is also $O(cn^2(m+2n))$. Thus, omitting a constant coefficient, the complexity of classical computing part of algorithm FindImpoDiff is 
\begin{equation*}
    \begin{split} 
&\sum_{r_1=1,\cdots,r-1\atop r_2=r-r_1} [cn^2(m+2n)+cn^2(m+2n)]\\
= &2c(r-1)n^2(m+2n)\in O(n^3).
    \end{split}
\end{equation*}
Therefore classical computing part has a complexity of $O(n^3)$.

\section{Improved Quantum Tool for Finding Impossible Differentials}

For some block ciphers, it is almost impossible to find a differential of which the probability is strictly 1. Thus many of attacks that have been proposed consider the truncated probability-1 differentials. For truncated differentials, only partial bits instead of the full differential are certain. For many block ciphers, such as SAFERK64 and Camellia, truncated differential analysis can attack more rounds than traditional differential analysis, or the attack complexity is greatly reduced when attacking the same number of rounds. In this section, we improve algorithm FindImpoDiff by allowing the differentials with probability 1 to be truncated differentials.

\subsection{Improved Algorithm for Impossible Differentials}

To improve algorithm FindImpoDiff, we allow the unmatched probability-1 differentials to be truncated differentials when applying miss-in-the-middle method. That is, only partial bits of the probability-1 differentials are predicted. 

Let $(\alpha,\beta)$ denote a truncated differential of $E^{(r)}:\mathcal{K}\times\mathbb{F}_2^n\rightarrow\mathbb{F}_2^n$, where $\alpha,\beta\in\{0,1,*\}^n$. Suppose $\beta=(\beta_1,\beta_2,\cdots,\beta_n)$, $\alpha=(\alpha_1,\alpha_2,\cdots,\alpha_n)$, then $\beta_i,\alpha_i,\in\{0,1,*\}$ for $i=1,\cdots,n$. The notation ``$*$'' indicates that the corresponding bits of the input or output difference  are undetermined. If $\alpha_i$ /$\beta_i$ $\in\{0,1\}$, then we call the $i$-th bit as a determined bit of $\alpha$ /$\beta$, otherwise, we call it an undetermined bit of $\alpha$ /$\beta$. A truncated  difference can actually be regarded as a set of full differences. Let 
$$
\Omega_{\alpha}=\big\{\Delta x=(\Delta x_1,\Delta x_2,\cdots,\Delta x_n)\in\mathbb{F}_2^n\big|\Delta x_i=\alpha_i \text{ if } \alpha_i\neq *, i=1,2,\cdots,n\big\},
$$
$$
\Omega_{\beta}=\Big\{\Delta y=(\Delta y_1,\Delta y_2,\cdots,\Delta y_n)\in\mathbb{F}_2^n\Big|\Delta y_i=\beta_i \text{ if } \beta_i\neq *, i=1,2,\cdots,n\Big\}.
$$
Then truncated input difference $\alpha$ is equivalent to the set $\Omega_{\alpha}$, and the truncated output difference $\beta$ is equivalent to the set $\Omega_{\beta}$. If a full input difference $\Delta x=(\Delta x_1,\cdots,\Delta x_n)\in\mathbb{F}_2^n$ is in the set $\Omega_{\alpha}$, that is, $\Delta x_i=\alpha_i$ for all $i\in\{1,\cdots,n\}$ such that $\alpha_i\neq *$, $\Delta x$ is said to coincide with the truncated input difference $\alpha$, and denote as $\Delta x \sim \alpha$. Similarly, if a full output difference  $\Delta y$ is in the set $\Omega_{\beta}$, we say that $\Delta y$ coincides with the truncated output difference $\beta$, and denote as $\Delta y \sim \beta$. Two truncated differentials $\alpha,\alpha'\in\{0,1,*\}^n$ are said to be contradict, if there is $i\in\{1,...,n\}$ satisfying $\alpha_i\neq*$, $\alpha'_i\neq*$ and $\alpha_i\neq\alpha'_i$. 

The probability of the truncated differential $(\alpha,\beta)$ is the conditional probability
\begin{align*}
\Pr_{k\leftarrow\mathcal{K}\atop x\leftarrow\mathbb{F}_2^n}[\alpha\overset{E^{(r)}}{\rightarrow}\beta]&=\Pr_{k\leftarrow\mathcal{K}\atop x\leftarrow\mathbb{F}_2^n}[E_k^{(r)}(x\oplus \Delta x)\oplus E_k^{(r)}(x)\sim \beta | \Delta x\sim\alpha ]\\
&=\Pr_{k\leftarrow\mathcal{K}\atop x\leftarrow\mathbb{F}_2^n}[E_k^{(r)}(x\oplus \Delta x)\oplus E_k^{(r)}(x)\in \Omega_{\beta}|\Delta x\in\Omega_{\alpha}].
\end{align*}
If this differential probability is equal to 1, we call $(\alpha,\beta)$ as a probability-1 truncated differential of $E^{(r)}$.

\begin{figure}[H]
\centering
\includegraphics[width=9cm]{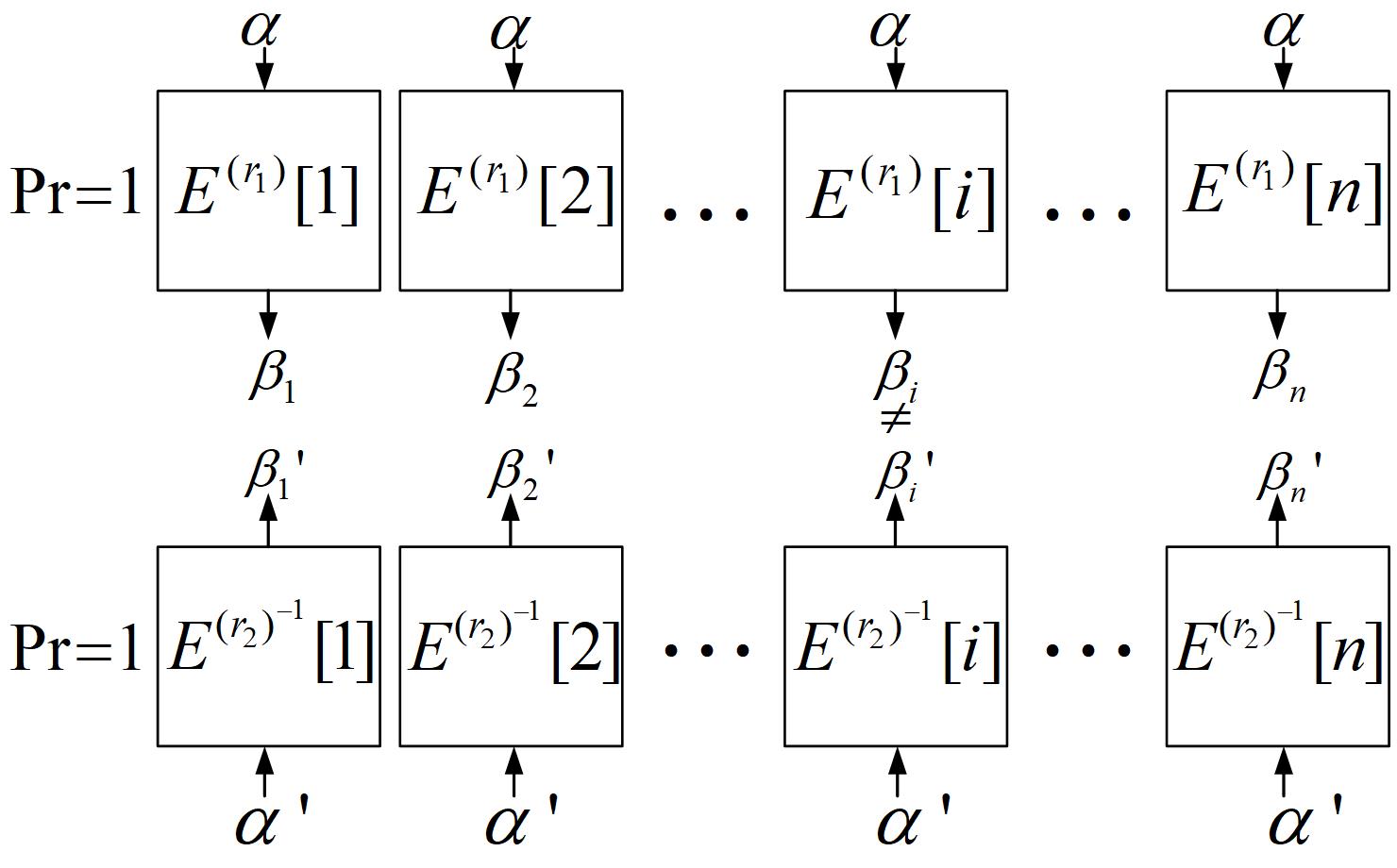}
\caption{Illustration of how truncated differentials constitute an impossible differential\label{Fig4}}
\end{figure}

We still divide $E^{(r)}=E^{(r_2)}\circ E^{(r_1)}$, where $r_1+r_2=r$. Let $E^{(r_1)}[i]$ be the $i$-th component function of $E^{(r_1)}$. That is, 
\begin{align*}
E_{k}^{(r_1)}(x)=(E_{k}^{(r_1)}[1](x),E_{k}^{(r_1)}[2](x),\cdots,E_{k}^{(r_1)}[n](x)).
\end{align*}
Similarly, ${E^{(r_2)}}^{-1}[i]$ denotes the $i$-th component function of ${E^{(r_2)}}^{-1}$. If the truncated differential $(\alpha,\beta)$ of $E^{(r_1)}$ has a probability of 1, the truncated differential $(\alpha',\beta')$ of ${E^{(r_2)}}^{-1}$ also has a probability of 1, and $\beta$ is contradict to $\beta'$, then $(\alpha,\alpha')$ will be $E^{(r)}$'s impossible differential. These conditions mean that, there exist $i\in\{1,\cdots,n\}$ such that $(\alpha,\beta_i)$ and $(\alpha',\beta'_i)$ are probability-1 differentials of $E^{(r_1)}[i]$ and ${E^{(r_2)}}^{-1}[i]$ respectively and $\beta_i\neq\beta'_i$ ($\beta_i,\beta'_i\neq*$). In this case $(\alpha,\alpha')$ will be $E^{(r)}$'s impossible differential, as shown in Figure~\ref{Fig4}. Thus, we only need to traverse $i$ to find the differentials of $E^{(r_1)}[i]$ and ${E^{(r_2)}}^{-1}[i]$ with probability 1, this can be done by using algorithm FindStruct to find their linear structures. According to these analysis, we design an improved algorithm that finds impossible differentials, which is named FindImpoDiff2.

\begin{algorithm}[H]
\caption*{{\large \textbf{Algorithm FindImpoDiff2}}}
\hangafter 1
\hangindent 3.7em
\textbf{Input$\,$}: $\,\,$a parameter $c$ and a block cipher $E^{(r)}:\mathcal{K}\times\mathbb{F}_2^n\rightarrow\mathbb{F}_2^n$ where $\mathcal{K}=\mathbb{F}_2^m$.

\hangafter 1
\hangindent 3.7em
\textbf{Output}: Impossible differentials of $E^{(r)}$.

\begin{algorithmic}[1]
    \For{$r_1=1$ to $r-1$}
        \State \parbox[t]{\dimexpr\textwidth-\leftmargin-\labelsep-\labelwidth}{Let $r_2=r-r_1$, divide $E^{(r)}$ into two parts $E^{(r)}=E^{(r_2)}\circ E^{(r_1)}$.}
        \For{$i=1$ to $n$\strut}
        \State \parbox[t]{\dimexpr\textwidth-\leftmargin-\labelsep-\labelwidth}{Run steps 1-7 of algorithm FindStruct on $E^{(r_1)}[i]:\mathbb{F}_2^{m}\times\mathbb{F}_2^n\rightarrow\mathbb{F}_2$ with parameter $c$, getting $c(1+n+m)$ vectors $\gamma^{(1)},\gamma^{(2)},\cdots,\gamma^{(c(1+n+m))}\in\mathbb{F}_2^{m}\times{F}_2^{n}\times{F}_2^{1}$.\strut}
        \State \parbox[t]{\dimexpr\textwidth-\leftmargin-\labelsep-\labelwidth}{Solve the equation
          \begin{equation*}\left\{\begin{array}{l} \gamma^{(1)}\cdot(k,x,y)=0\\ \gamma^{(2)}\cdot(k,x,y)=0\\\vdots\\ \gamma^{(c(1+n+m))}\cdot(k,x,y)=0\\
          k=0^{m},
          \end{array}\right.\end{equation*}
          where $(k,x,y)\in\mathbb{F}_2^{m+n+1}$ are unknowns, to get the solution set $A_{r_1}^i$. \strut}
        \State \parbox[t]{\dimexpr\textwidth-\leftmargin-\labelsep-\labelwidth}{Run steps 1-7 of algorithm FindStruct on ${E^{(r_2)}}^{-1}[i]:\mathbb{F}_2^{m}\times\mathbb{F}_2^n\rightarrow\mathbb{F}_2$ with parameter $c$, getting $c(1+n+m)$ vectors $\tilde{\gamma}^{(1)},\tilde{\gamma}^{(2)},\cdots,\tilde{\gamma}^{(c(1+n+m))}\in\mathbb{F}_2^{m+n+1}$.\strut}
\algstore{bkbreak}
\end{algorithmic}
\end{algorithm}

\begin{algorithm}
\begin{algorithmic}[1]
\algrestore{bkbreak}
        \State \parbox[t]{\dimexpr\textwidth-\leftmargin-\labelsep-\labelwidth}{Solve the equation
          \begin{equation*}\left\{\begin{array}{l} \tilde{\gamma}^{(1)}\cdot(k,x,y)=0\\ \tilde{\gamma}^{(2)}\cdot(k,x,y)=0\\\vdots\\\tilde{\gamma}^{(c(1+n+m))}\cdot(k,x,y)=0\\
          k=0^{m},
          \end{array}\right.\end{equation*}where $(k,x,y)\in\mathbb{F}_2^{m+n+1}$ are unknowns, to get the solution set $B_{r_2}^i$. \strut}  
        \For{$(0^{m},\Delta x_1,\Delta y_1)\in A_{r_1}^i$}
        \For{$(0^{m},\Delta x_2,\Delta y_2)\in B_{r_2}^i$}
        \If{$\Delta x_1\neq 0^n$  $\wedge$ $\Delta x_2\neq 0^n$  $\wedge$  $\Delta y_1\neq \Delta y_2$}
        \State \parbox[t]{\dimexpr\textwidth-\leftmargin-\labelsep-\labelwidth}{Output $(\Delta x_1,\Delta x_2)$.\strut}
        \EndIf
         \EndFor
         \EndFor
         \EndFor
    \EndFor
\end{algorithmic}
\end{algorithm}

Algorithm FindImpoDiff2 traverses $r_1$ from 1 to $r-1$, dividing $E^{(r)}$ into $E^{(r)}=E^{(r_2)}\circ E^{(r_1)}$, and then traverses $i$ from 1 to $n$ to obtain differentials of $E^{(r_1)}[i]$ and ${E^{(r_2)}}^{-1}[i]$ that have probability of 1 but lead to contradiction in the middle. Define the parameter
\begin{align*}
    \bar{\Theta}(E^{(r)})=\max\{\theta(E^{(t)}[i])|1\leq t\leq r-1, 1\leq i\leq n\},
\end{align*}
where $E^{(t)}$ is the $t$-round reduced cipher of $E^{(r)}$, $E^{(t)}[i]$ is the $i$-th component function of $E^{(t)}$, $\theta(E^{(t)}[i])$ is defined as Eq.(6). That is,
\begin{equation*}
    \begin{split}        \theta(E^{(t)}[i])=\max_{{(a_1,a_2)\in\mathbb{F}_2^{m}\times\mathbb{F}_2^{n}\atop b=0,1}\atop ((a_1,a_2),b)\notin L_{E^{(t)}[i]}}\frac{1}{2^{m+n}}\Big|\big\{(k,x)\in\mathbb{F}_2^{m}\times\mathbb{F}_2^{n}|E^{(t)}[i](k,x)\oplus\\
    E^{(t)}[i](k\oplus a_1,x\oplus a_2)=b\big\}\Big|,
    \end{split}
\end{equation*}
where $L_{E^{(t)}[i]}$ is the linear structure space of $E^{(t)}[i]$. The following theorem holds according to Theorem 2.
\begin{Theorem} $E^{(r)}$ is a block cipher whose round number is $r$, blocksize is $n$ and key length is $m$. $\bar{\Theta}(E^{(r)})\leq e_0<1$ for constant $e_0$. If $(\Delta x_1,\Delta x_2)$ is output by algorithm FindImpoDiff2 when running on $E^{(r)}$, then the probability of $(\Delta x_1,\Delta x_2)$ being an impossible differential of $E^{(r)}$is no less than $1-2\big(2(\frac{1+e_0}{2})^c\big)^{m+n+1}$.
\end{Theorem}

According to Theorem 4, taking $c$ greater than $3/(1-e_0)$ guarantees that algorithm FindImpoDiff2 outputs impossible differentials of $E^{(r)}$.

\subsection{Complexity of Algorithm FindImpoDiff2}
The process of algorithm FindImpoDiff does not involve quantum queries and thus can be executed by Q1 adversaries. We still evaluate the complexity by calculating the amounts of qubits and quantum gates.

In algorithm FindImpoDiff2, finding probability-1 differentials of $E^{(r_1)}[i]$ requires to execute 1-7 steps of algorithm \textbf{FindStruct} on $E^{(r_1)}[i]$ with parameter $c$. This needs $2c(1+n+m)^2$ Hadamard gates, $c(1+n+m)$ CNOT gates and $c(1+n+m)$ executions of unitary operator $U_{E^{(r_1)}[i]}$. Similarly, finding probability-1 differentials of ${E^{(r_2)}}^{-1}$ needs $2c(1+n+m)^2$ Hadamard gates, $c(1+n+m)$ CNOT gates and $c(1+n+m)$ executions of unitary operator $U_{{E^{(r_2)}}^{-1}[i]}$. Thus, the amount of Hardamard gates in algorithm FindImpoDiff2 is
\begin{equation*}
    \begin{split} 
&\sum_{r_1=1,\cdots,r-1\atop r_2=r-r_1}\sum_{i=1}^n[2c(1+n+m)^2+2c(1+n+m)^2]\\
&=\sum_{r_1=1}^{r-1}\sum_{i=1}^n[4c(1+n+m)^2]\\
&=4(r-1)cn(1+n+m)^2\in O(n^3).
    \end{split}
\end{equation*}
The amount of CNOT gates in algorithm FindImpoDiff2 is
\begin{equation*}
    \begin{split} 
&\sum_{r_1=1,\cdots,r-1\atop r_2=r-r_1}\sum_{i=1}^n[c(1+n+m)+c(1+n+m)]\\
&=2(r-1)cn(1+n+m)\in O(n^2).
    \end{split}
\end{equation*}
Algorithm FindImpoDiff2 also requires to execute the unitary operators $U_{E^{(r_1)[i]}}$ and $U_{{E^{(r_2)}}^{-1}[i]}$ $c(m+n+1)$ times for each $1\leq r_1\leq r-1$, $r_2=r-r_1$ and $1\leq i\leq n$. Since 
\begin{equation*}
    \begin{split} 
&\sum_{r_1=1,\cdots,r-1\atop r_2=r-r_1}\sum_{i=1}^n\Big[c(m+n+1)|U_{E^{(r_1)}[i]}|+c(1+n+m)|U_{{E^{(r_2)}}^{-1}[i]}|\Big]\\
&= c(1+n+m)\sum_{r_1=1,\cdots,r-1\atop r_2=r-r_1}\sum_{i=1}^n\Big(|U_{E^{(r_1)}[i]}|+\sum_{i=1}^n|U_{{E^{(r_2)}}^{-1}[i]}|\Big)\\
&= c(1+n+m)\sum_{r_1=1,\cdots,r-1\atop r_2=r-r_1}\big(|U_{E^{(r_1)}}|+|U_{E^{(r_2)}}|\big)\\
&=(r-1)c(1+n+m)|U_{E^{(r)}}|\in O(poly(n)),
    \end{split}
\end{equation*}
algorithm FindImpoDiff2 additionally requires to execute the quantum circuit of $E^{(r)}$ for $(r-1)c(1+n+m)$ times. The quantum resources to implement algorithm FindImpoDiff2 are listed in Tabla 1.

Running FindStruct subroutine on $E^{(r_1)}[i]$ needs $(m+n)+1+1=m+n+2$ qubits. Running FindStruct subroutine on ${E^{(r_2)}}^{-1}[i]$ also needs $m+n+2$ qubits. Due to the reusability of qubits, $m+n+2$ qubits are enough for algorithm FindImpoDiff2.

\section{Results and Discussion}

In this work, we develop quantum automatic cryptanalysis tools for searching impossible differentials. Our tools combine the impossible differential attack with Simon’s algorithm. We rigorously prove that, if an impossible differential of the block cipher is linked by two truncated differentials with probability 1, the proposed quantum algorithms must be able to output these impossible differentials in polynomial time. To analyze the complexity, we calculated the number of various quantum operators required in the quantum algorithms and summarized them in Table 1. 

Our quantum cryptanalytic tools do no need any query to the encryption or decryption oracle and can be implemented in Q1 model. Thus they are easy to realize with quantum computers. Compared to the classical automatic impossible differential cryptanalysis, our quantum automatic tools fully make use of the superiority of quantum computing, allowing for detailed characterization of S-boxes while maintaining complexity within polynomial time. Moreover, the proposed quantum cryptanalytic tools take the key schedule into account in single-key model, making up for the shortcomings of traditional tools. 

The natural directions for further research is to improved the cryptanalytic tools proposed in this paper to achieve less consumption of quantum resources, or enhance other cryptanalytic methods with quantum algorithms, such as integral and linear attacks. Combining cryptanalytic tools with Grover's algorithm may also be a direction worth exploring. 

\begin{table}[H] 
    \caption{The quantum resources of algorithm FindImpoDiff and algorithm FindImpoDiff2\textsuperscript{1}}
\begin{tabular}{cccc}
\hline
\textbf{Algorithm}	& \textbf{\#CNOT}	& \textbf{\#Hadamard} & \textbf{$U_{E^{(r)}}$}\\
\hline
\textbf{FindImpoDiff}		& $2\tau (2n^2+nm)$ & $4\tau (m^2+4n^2+4mn)$			& $\tau (m+2n)$\\
\textbf{FindImpoDiff2}		& $2\tau (n^2+nm+n)$ & $4\tau n(n+m+1)^2$			& $\tau (1+n+m)$\\
\hline
\end{tabular}
\noindent{\footnotesize{\textsuperscript{1} Here $n$, $m$ and $r$ is the blocksize, length of key and the amount of rounds, respectively. $c$ is the parameter chosen by the attacker}, $\tau=c\cdot(r-1)$. }
\end{table}

\section*{Acknowledgement}
This research was supported by Beijing Natural Science Foundation (no.4234084) and the Open Research Fund of Key Laboratory of Cryptography of Zhejiang Province (no. ZCL21012).

\begin{appendix}
\section{}
\noindent
\textbf{Theorem 2} \textit{Let $L$ be the solution set output by algorithm FindStruct run on $F:\mathbb{F}_2^n\rightarrow\mathbb{F}_2^m$ with a parameter $c$, then $L_F\subseteq L$. Moreover, if there is a constant $e_0$ such that $\theta(F)\leq e_0<1$, then the probability of $L_F$ being equal to $L$ is no less than $1-\big(2(\frac{1+e_0}{2})^c\big)^{n+m}$.}
\vspace{0.2cm}

\begin{proof}[Proof]
First prove that $L_F\subseteq L$. To do this, we only need to prove that, each linear structure $(a,b)\in L_F$ must be a solution of Eq.(5). In the step 5 of algorithm FindStruct, the state
$$|\Psi_3\rangle=\frac{1}{\sqrt{2^{n+m}}}\sum_{x\in\mathbb{F}_2^n, y\in\mathbb{F}_2^m}|x\rangle|y\rangle|F(x)\oplus y\rangle$$
is measured and the measurement result is denoted by $z$. Then there are $x_0\in\mathbb{F}_2^n$, $y_0\in\mathbb{F}_2^m$ such that $F(x_0)\oplus y_0=z$. Define the set
$$
S_z=\big\{(x,y)\in\mathbb{F}_2^n\times\mathbb{F}_2^m|F(x)\oplus y=z\big\},
$$
then the first two registers of $|\Psi_3\rangle$ are collapsed to the state $\frac{1}{\sqrt{|S_z|}}\sum_{(x,y)\in S_z}|x\rangle|y\rangle$. Obviously $(x_0,y_0)\in S_z$. For any $(x,y)\in\mathbb{F}_2^{n+m}$, let $a=x_0\oplus x, b=y_0\oplus y$, then $y=b\oplus y_0$, $x=a\oplus x_0$ and
\begin{align*}
&F(x)\oplus y=z\\
\Leftrightarrow&(b\oplus y_0)\oplus F(a\oplus x_0)=z\\
\Leftrightarrow& F(a\oplus x_0)\oplus b\oplus y_0=y_0\oplus F(x_0)\\
\Leftrightarrow& F(x_0)\oplus F(x_0\oplus a)=b\\
\Leftrightarrow& (a,b) \text{ causes a \emph{match} of }F\text{ at point }x_0.\\
\end{align*}
Thus 
\begin{align*}
S_z=\big\{(x_0\oplus a,y_0\oplus b)\big|(a,b)\in\mathbb{F}_2^{n+m}\text{ and }(a,b)
\text{ causes a \emph{match} of } F \text{ at } x_0\}.
\end{align*}
Since the linear structure duads of $F$ cause \emph{matches} of $F$ at all points $x\in\mathbb{F}_2^n$, for each $(a,b)\in L_F$, $(x_0\oplus a,y_0\oplus b)$ is in $S_z$. Suppose that $\{(a_1,b_1),(a_2,b_2),\cdots,$ $(a_t,b_t)\}$ is the basis of space $L_F$, then for any $k_1,k_2,\cdots,k_t\in\{0,1\}$, we have 
$$
(k_1a_1\oplus k_2a_2\oplus...\oplus k_ta_t,k_1b_1\oplus k_2b_2\oplus...\oplus k_tb_t)\in L_F.
$$
Therefore
$$(x_0\oplus\bigoplus_{j=1}^tk_ja_j,y_0\oplus\bigoplus_{j=1}^tk_jb_j)\in S_z.$$
In addition to linear structures, there may be other vectors causing \emph{match} at $x_0$. Let $(\hat{a},\hat{b})$ denote such vectors. Namely, $F(x_0\oplus\hat{a})\oplus(y_0\oplus\hat{b})=z$ but $(\hat{a},\hat{b})\notin L_F$. Thus for any $k_1,\cdots,k_t\in\{0,1\}$, 
\begin{align*}
F(x_0\oplus\hat{a}\oplus\bigoplus_{j=1}^tk_ja_j)\oplus(y_0\oplus\hat{b}\oplus\bigoplus_{j=1}^tk_jb_j)
=F(x_0\oplus \hat{a})\oplus(y_0\oplus \hat{b})=z
\end{align*}
So $(x_0\oplus\hat{a}\oplus\bigoplus_{j=1}^tk_ja_j,y_0\oplus\hat{b}\oplus\bigoplus_{j=1}^tk_jb_j)$ is also in the set $S_z$. Suppose $(\hat{a}_1,\hat{b}_1), (\hat{a}_2,\hat{b}_2)$ are two such vectors, i.e., they both cause \emph{match} at $x_0$ but are not in $L_F$. Since $L_F$ is a linear space, the following two sets 
$$
\big\{(x_0\oplus\hat{a}_1\oplus \bigoplus_{j=1}^tk_ja_j,y_0\oplus \hat{b}_1\oplus \bigoplus_{j=1}^tk_jb_j)\big|k_1\cdots,k_t\in\!\!\{0,1\}\big\} 
$$
$$
\big\{(x_0\oplus\hat{a}_2\oplus \bigoplus_{j=1}^tk_ja_j,y_0\oplus \hat{b}_2\oplus\bigoplus_{j=1}^tk_jb_j)\big|k_1,\cdots,k_t\in\!\!\{0,1\}\big\}
$$
are either equal or have no intersection at all. Therefore, the measurement in step 5 of algorithm FindStruct causes the first two registers of $|\Psi_3\rangle$ collapsed to a state with the following form:
\begin{equation*}
\begin{split}
&\frac{1}{\sqrt{2^t(l+1)}}\,\,\,\,\,\Big(\,\,\sum_{\mathclap{k_1,\cdots,k_t\in\{0,1\}}}\,\,\,|x_0\oplus \bigoplus_{j=1}^tk_ja_j\rangle|y_0\oplus \bigoplus_{j=1}^tk_jb_j\rangle\\
&+\,\,\,\sum_{\mathclap{k_1,\cdots,k_t\in\{0,1\}}}\,\,\,|x_0\oplus\hat{a}_1\oplus \bigoplus_{j=1}^tk_ja_j\rangle|y_0\oplus \hat{b}_1\oplus \bigoplus_{j=1}^tk_jb_j\rangle\\
&+\cdots\\
&+\,\,\,\sum_{\mathclap{k_1,\cdots,k_t\in\{0,1\}}}\,\,\,|x_0\oplus\hat{a}_l\oplus \bigoplus_{j=1}^tk_ja_j\rangle|y_0\oplus \hat{b}_l\oplus \bigoplus_{j=1}^tk_jb_j\rangle\Big),
\end{split}
\end{equation*}
where $(\hat{a}_1,\hat{b}_1),\cdots,(\hat{a}_l,\hat{b}_l)$ are the vectors that cause \emph{matches} at $x_0$ but not in $L_F$. Denote $\hat{a}_0=0^n$, $\hat{b}_0=0^m$, this state can be written as
$$
\frac{1}{\sqrt{2^t(l+1)}}\!\!\sum_{i\in\{0,1,\cdots,l\}\atop k_1,\cdots,k_t\in\{0,1\}}\!\!\!|x_0\oplus \hat{a}_i\oplus\bigoplus_{j=1}^tk_ja_j\rangle|y_0\oplus\hat{b}_i\oplus \bigoplus_{j=1}^tk_jb_j\rangle.
$$
In the step 6, after performing Hadamard gate $H^{\otimes(n+m)}$, the first two registers become
\begin{equation*}
\begin{split}
&\,\,\,\,\,\,\,\sum_{\mathclap{i\in\{0,\cdots,l\} \atop {k_1,\cdots,k_t\in\{0,1\}\atop (\gamma_1,\gamma_2)\in\mathbb{F}_2^{n+m}}}}(-1)^{\gamma_1\cdot(x_0\oplus \hat{a}_i\oplus\bigoplus_{j=1}^tk_ja_j)+\gamma_2\cdot(y_0\oplus\hat{b}_i\oplus \bigoplus_{j=1}^tk_jb_j)}|\gamma_1\rangle|\gamma_2\rangle\\
&=\!\!\sum_{\mathclap{\gamma_1\in\mathbb{F}_2^{n}\atop \gamma_2\in\mathbb{F}_2^{m}}}\big(\sum_{i=0}^l(-1)^{\gamma_1\cdot (x_0\oplus\hat{a}_i)\oplus \gamma_2\cdot (y_0\oplus\hat{b}_i)}\big)\big[(-1)^{(\gamma_1,\gamma_2)\cdot(a_1,b_1)}+1\big]\times\\
&\quad\qquad\big[(-1)^{(\gamma_1,\gamma_2)\cdot(a_2,b_2)}+1\big]\times\cdots\times\big[(-1)^{(\gamma_1,\gamma_2)\cdot(a_t,b_t)}+1\big]|\gamma_1,\gamma_2\rangle,
\end{split}
\end{equation*}
where we omit the global coefficient $1/\sqrt{2^{n+m+t}(l+1)}$. For any $(\gamma_1,\gamma_2)\in\mathbb{F}_2^{n+m}$, if there exists $j\in\{1,2,\cdots,t\}$ such that $(\gamma_1,\gamma_2)\cdot(a_j,b_j)\neq0$, then the amplitude of $|\gamma_1,\gamma_2\rangle$ in the above quantum state is zero. Therefore, algorithm FindStruct always outputs a random value $\gamma\in\mathbb{F}_2^{n+m}$ such that $\gamma\cdot(a_j,b_j)=0$ for all $j=1,2,\cdots,t$, which means $(a_1,b_1),\cdots,(a_t,b_t)$ must be in the solution space $L$ of Eq.(5). Thus we have $L_F\subseteq L$.

Then prove the probability that $L_F=L$ to be no less than $1-2((\frac{1+e_0}{2})^c)^{n+m}$ as long as $\theta(F)\leq e_0<1$. $L_F\neq{L}$ means that there is a vector $(a,b)$ which is a solution of Eq.(5) but not in $L_F$. Thus, 
\begin{align}
&{\rm Pr}\big[L_F\neq L]\notag \\\notag
=&{\rm Pr}\big[\exists (a,b)\notin L_F, s.t.,\,\,\gamma^{(1)}\cdot (a,b)=\gamma^{(2)}\cdot (a,b)=\cdots =\gamma^{(c(n+m))}\cdot (a,b)=0\,\big]\notag \\\notag
\leq&\,\,\,\,\,\,\sum_{\mathclap{(a,b)\in\mathbb{F}_2^{n+m}\backslash L_F}}\,\,\,{\rm Pr}\big[\gamma^{(1)}\cdot (a,b)=\cdots =\gamma^{(c(n+m))}\cdot (a,b)=0\big]\\\notag
=&\sum_{(a,b)\in\mathbb{F}_2^{n+m}\backslash L_F}\!\!\!\big(\,{\rm Pr}[\gamma^{(1)}\cdot (a,b)=0]\,\big)^{c(n+m)}\\\notag
\leq&(2^{n+m}-|L_F|)\,\,\,\,\,\max_{\mathclap{(a,b)\in\mathbb{F}_2^{n+m}\backslash L_F}}\,\,\,\,\,\big({\rm Pr}[\gamma^{(1)}\cdot (a,b)=0]\big)^{c(n+m)}\\
\leq&\max_{(a,b)\in\mathbb{F}_2^{n+m}\backslash L_F}\Big(2{\rm Pr}[\gamma^{(1)}\cdot (a,b)=0]^c\Big)^{n+m},
\end{align}
where $\gamma^{(1)},\cdots,\gamma^{(c(n+m))}$ are $c(m+n)$ outputs of FindStruct subroutine and are independent and identically distributed  random variables. To calculate the probability ${\rm Pr}[\gamma^{(1)}\cdot (a,b)=0]$, all measurements of FindStruct subroutine are moved to the end. According to principle of deferred measurement, this does not change the outputs. Therefore the state without being measured is 
\begin{equation*}
\begin{split}
&\frac{1}{2^{n+m}}\sum_{x\in\mathbb{F}_2^n\atop y\in\mathbb{F}_2^m}\sum_{\gamma_1\in\mathbb{F}_2^n\atop \gamma_2\in\mathbb{F}_2^m}(-1)^{x\cdot \gamma_1\oplus y\cdot \gamma_2}|\gamma_1\rangle|\gamma_2\rangle|F(x)\oplus y\rangle\\
=&\frac{1}{2^{n+m}}\sum_{(\gamma_1,\gamma_2)\in\mathbb{F}_2^{n+m}\atop (\gamma_1,\gamma_2)\cdot(a,b)=0}\sum_{x\in\mathbb{F}_2^n\atop y\in\mathbb{F}_2^m}(-1)^{x\cdot \gamma_1\oplus y\cdot \gamma_2}|\gamma_1\rangle|\gamma_2\rangle|F(x)\oplus y\rangle\\
&+\frac{1}{2^{n+m}}\!\!\!\sum_{(\gamma_1,\gamma_2)\in\mathbb{F}_2^{n+m}\atop (\gamma_1,\gamma_2)\cdot(a,b)=1}\sum_{x\in\mathbb{F}_2^n\atop y\in\mathbb{F}_2^m}(-1)^{x\cdot \gamma_1\oplus y\cdot \gamma_2}|\gamma_1\rangle|\gamma_2\rangle|F(x)\oplus y\rangle.
\end{split}
\end{equation*}
The probability of $\gamma^{(1)}$ satisfying $(a,b)\cdot\gamma^{(1)}=0$ is 
\begin{equation*}
\begin{split}
&{\rm Pr}[(a,b)\cdot \gamma^{(1)}=0]\\
=&\big\|\frac{1}{2^{n+m}}\sum_{(\gamma_1,\gamma_2)\in\mathbb{F}_2^{n+m}\atop (\gamma_1,\gamma_2)\cdot(a,b)=0}\sum_{x\in\mathbb{F}_2^n\atop y\in\mathbb{F}_2^m}(-1)^{x\cdot \gamma_1\oplus y\cdot \gamma_2}|\gamma_1\rangle|\gamma_2\rangle|F(x)\oplus y\rangle\big\|^2 \notag\\
=&\frac{1}{2^{2(n+m)}}\,\,\sum_{\mathclap{(\gamma_1,\gamma_2)\cdot(a,b)=0\atop{x,x'\in\mathbb{F}_2^n \atop {y,y'\in\mathbb{F}_2^m}}}}(-1)^{\gamma_1\cdot (x\oplus x')\oplus \gamma_2\cdot(y\oplus y')}\langle F(x')\oplus y'|F(x)\oplus y\rangle\\
=&\,\,\,\sum_{\mathclap{x,x'\in\mathbb{F}_2^n \atop {y,y'\in\mathbb{F}_2^m}}}\Big(\frac{\langle y\oplus F(x)| y'\oplus F(x')\rangle}{2^{2(n+m)}}\,\,\sum_{\mathclap{(\gamma_1,\gamma_2)\cdot(a,b)=0}}(-1)^{\gamma_1\cdot (x'\oplus x)\oplus \gamma_2\cdot(y'\oplus y)}\Big).
\end{split}
\end{equation*}
According to the Lemma 1 in \cite{ref-proceeding5}, it holds that
$$
\frac{1}{2^{n+m}}\sum_{(\gamma_1,\gamma_2)\in\mathbb{F}_2^{n+m}\atop (\gamma_1,\gamma_2)\cdot(a,b)=0}(-1)^{\gamma_1\cdot x+\gamma_2\cdot y}=\frac{1}{2}(\delta_{(0^n,0^m),(x,y)}+\delta_{(a,b),(x,y)}).
$$
Thus
\begin{equation}
\begin{split}
&{\rm Pr}[(a,b)\cdot\gamma^{(1)}=0]\\
=&\,\,\,\,\,\sum_{\mathclap{(x,y)\in\mathbb{F}_2^{n+m}\atop (x',y')\in\mathbb{F}_2^{n+m}}}\,\,\,\frac{\langle y\oplus F(x)| y'\oplus F(x')\rangle}{2^{n+m+1}}\Big(\delta_{(x,y),(x',y')}+\delta_{(x,y),(a\oplus x',b\oplus y')}\Big)\notag.\\
=&\frac{1}{2^{n+m+1}}\sum_{(x,y)\in\mathbb{F}_2^{n+m}}\Big(1+\langle F(x)\oplus y|F(x\oplus a)\oplus y\oplus b\rangle\Big) \notag.\\
=&\frac{1}{2}\big(1+\frac{1}{2^{n+m}}|\{(x,y)|F(x\oplus a)\oplus F(x)=b\}|\big)\\
=&\frac{1}{2}\big(1+{\rm Pr}_x[F(x\oplus a)\oplus F(x)=b]\big)\\
\leq&\frac{1}{2}\Big(1+\theta(F))\Big).\\
\end{split}
\end{equation}
Combining Eq.(A1) we have
\begin{align*}
&{\rm Pr}[L_F\neq L]\\
\leq&\max_{(a,b)\in\mathbb{F}_2^{m+n}\backslash L_F}\Big(2{\rm Pr}[(a,b)\cdot\gamma^{(1)}=0]^c\Big)^{n+n}\\
\leq&\Big(2(\frac{1+\theta(F)}{2})^c\Big)^{m+n}\\
\leq&\Big(2(\frac{1+e_0}{2})^c\Big)^{m+n}.
\end{align*}
\end{proof}
\end{appendix}
\end{document}